\documentclass[manuscript]{aastex}

\usepackage{graphicx}
\usepackage{amssymb,amsmath}
\usepackage[english]{babel}
\usepackage{url}
\usepackage{ulem}
\usepackage{xcolor}

\usepackage{epstopdf}

\begin{document}
	
	\title{THREE-DIMENSIONAL KINETIC-MHD MODEL OF THE GLOBAL HELIOSPHERE WITH THE HELIOPAUSE-SURFACE FITTING}
	
	\author{ V. V. Izmodenov\altaffilmark{1,2}, D.B. Alexashov\altaffilmark{2}}
	\affil{Space Research Institute of Russian Academy of Sciences, Moscow, Russia}
	
	\email{izmod@iki.rssi.ru}
	
	\altaffiltext{1}{Lomonosov Moscow State University, Moscow, Russia.}
	\altaffiltext{2}{Institute for Problems in Mechanics of Russian Academy of Sciences, Moscow, Russia.}

	\begin{abstract}
		
This paper provides a detailed description of the latest version of our model
of the solar wind (SW) interaction with the local interstellar medium
(LISM). This model has already been applied to the 
analysis of Lyman-alpha absorption spectra toward nearby stars and for analyses of 
{\it Solar and Heliospheric Observatory}/SWAN data. Katushkina et al. (this issue) used the model results to analyze {\it IBEX}-Lo data.  At the same time, the details of this model have not yet been published.
This is a three-dimensional (3D) kinetic-magnetohydrodynamical (MHD) model that takes into account SW and  interstellar plasmas (including $\alpha$ particles in SW
and helium ions in LISM), the solar and interstellar
magnetic fields, and the interstellar hydrogen atoms. The  latitudinal dependence of SW and the actual flow
direction of the interstellar gas with respect to the Sun are also taken into account in the model.
It was very essential that our numerical code  had been developed in such a way that any numerical diffusion or reconnection across the heliopause had not been allowed in the model.
The heliospheric current sheet is a rotational discontinuity in the ideal MHD and can be treated kinematically.
In the paper, we focus in particular on the effects of the heliospheric magnetic field and on the heliolatitudinal dependence of SW.

\end{abstract}



\bibliographystyle{aj}

\section{Introduction}

 Global modeling of the solar wind (SW) interaction with the local interstellar medium (LISM)  is essential: (1) in general, to understand how the Sun acts as a star and exchanges matter and energy with the LISM, and (2) in particular, to explain relevant space experiment data (e.g. obtained on board of both {\it Voyager} spacecraft, the {\it Interstellar Boundary Explorer} ({\it IBEX}), the {\it Solar and Heliospheric Observatory} ({\it SOHO}), the {\it Hubble Space Telescope} ({\it HST}), etc.).

Following the pioneering paper by Baranov \& Malama (1993) self-consistent kinetic-gas dynamics models of the SW/LISM interaction were developed by our group in Moscow (e.g. Izmodenov and Baranov, 2006, Izmodenov et al., 2009, for reviews). There are two features which make our model approach different from similar models developed by other groups. 
These features are as follow:
(1) a rigorous kinetic treatment of the interstellar H atom component using a Monte Carlo method with splitting of the trajectories,
and (2) a Godunov-type numerical method used to solve the ideal gas dynamical or MHD equations. The method has the potential to fit some major discontinuities of the flow and to capture the others.

The kinetic treatment of H atoms is essential because their mean free path is comparable to the size of the SW/LISM interaction region (see, e.g.,Izmodenov 2000). The difference between kinetic and alternative multi-component models has been explored in detail by Alexashov \& Izmodenov (2005) and M\"{u}ller et al. (2008). At the heliospheric distances smaller than 10-20 AU, solar gravitation and radiation forces make the kinetic effects even more important.
The velocity distribution function of H atoms is essentially non-Maxwellian (Izmodenov 2001; Izmodenov et al., 2013) in this region.  This is very important for analyses of any measurements (direct or indirect) of H atoms.

To obtain an effective tool to calculate the H atom space and velocity distribution inside the heliosphere, Katushkina \& Izmodenov (2010) suggested a two-step procedure. In the first step, the  H atom distribution is calculated in the frame of a  global self-consistent model of the SW/LISM interaction. As a particular result of this model, we obtain the velocity distribution of H atoms at a certain heliocentric distance (e.g., 90 AU)  where local solar effects could be neglected. This distribution function is used as a boundary condition for the second step in which the velocity distribution inside the heliosphere is calculated with the resolution (in time, space, and velocity space) required for the data analyses. The kinetic equation at this step is solved using the  method of characteristics in 3D time-dependent case (Izmodenov et al., 2013).

  The two-step procedure described above  has been used in the companion paper by Katushkina et al. (2015) in order to analyze the {\it IBEX}-Lo data for the interstellar hydrogen fluxes. The aim of this paper is to describe the global model of the SW/LISM interaction that has been employed in the companion paper during the first step of the two-step procedure. The model presented is a currently state-of-the-art model which we developed. It has not yet been described in the  literature.

 The original Baranov and Malama (1993) model is a stationary axisymmetric kinetic-gas dynamical model. This model has been advanced by considering additional physical components - galactic and anomalous cosmic rays (Myasnikov et al., 2000, Alexashov et al 2004), interstellar helium ions and SW alpha particles (Izmodenov et al., 2003), and interstellar oxygen and nitrogen (Izmodenov et al., 2004). This model has also been  extended in the tail direction  to distances of $\sim$20000-30000 AU (Izmodenov \& Alexashov 2003). In 2005, the model became  time-dependent (Izmodenov et al., 2005a), and then realistic SW data were used in the model (Izmodenov et al., 2008).  In 2005, the stationary version of this model was expanded to three dimensions and the interstellar magnetic field (IsMF) was taken into account (Izmodenov et al., 2005b). This model allowed us to explain quantitatively  the difference in the directions of the interstellar hydrogen and helium flows inside the heliosphere which had been discovered by Lallement et al. (2005, 2010). The model provided an asymmetry of the distances to the heliospheric termination shock (TS), in agreement with {\it Voyager 1} and {\it Voyager 2} data (Izmodenov, 2009). Vincent et al. (2014) used this model to analyze the backscattered Lyman-alpha spectra obtained by {\it HST}.

The Izmodenov et al. (2005b) model has been criticized because it did not take into account the (HMF) and variations of the SW parameters with the heliolatitude. The importance of HMF on the plasma flow in the inner heliosheath was realized very early by Axford et al. (1963) and Nerney \& Suess (1975), and later in the kinematic approach in the frame of the  Baranov and Malama model by Barsky et al. (1999). The dynamical effects of the HMF have been explored by other groups in the models of Washimi (1993), Linde et al. (1998), Pogorelov et al. (2004), Opher et al. (2004), and Ratkiewicz et al. (2006).
The heliolatitudinal variations of the SW on the global structure of the SW/LISM interaction region were previously studied by Tanaka \& Washimi (1999), Grygorczuk et al. (2010), Kim et al. (2012), Pogorelov et al. (2013), and Provornikova et al. (2014). More recent work on the SW/LISM modeling includes the studies of Heerikhuisen et al (2014), Zirnstein et al (2014), Zank et al. (2013), Ben-Jaffel \& Ratkiewicz (2012), Florinski et al (2013), Opher et al (2015), Borovikov \& Pogorelov (2014).

In this paper, our 3D kinetic-MHD model is extended by taking into account the dynamic effects of HMF in the inner heliosheath region, as well as the effect of the latitudinal variations of the SW parameters.
Note that the model presented here has already been used in a number of studies. For example, Wood et al. (2014) employed the model to simulate the Lyman-alpha absorption spectra in the tail of the heliosphere. Katushkina et al. (2015) used the model results to analyze the deflection of the direction of the H atom velocity vector in the heliosphere from its direction in the LISM. Finally, in the companion paper (Katushkina et al., 2015), this model has been used for the analysis of {\it IBEX}-Lo data.

Despite of the fact that the model has already been employed for  data analyses, a detailed description of the model had not been presented in the literature, although it had been reported at several conferences. The aim of this paper is to fill this gap and present a detailed formulation of the model (section 2) and its results (section 3). The effects of the HMF and latitudinal variations of the SW on the global structure of the SW/LISM interaction region will be explored explicitly by comparisons with models where the effects are not taken into account. To our knowledge, this has not yet been done. The relevance/irrelevance of the model results to observational data is discussed in section 4.

\section{Model}

In the model, partially ionized interstellar plasma is considered  as a two-component gas.
The two components are (1) a neutral component consisting of atomic hydrogen and (2) a charged (or plasma) component consisting of protons, electrons, and helium ions. In the SW, the plasma component consists of protons, alpha particles, and electrons.

The neutral component is described kinetically by means of a velocity distribution function,  $f_\mathrm{H} \left( \mathbf{r},
\mathbf{w}_\mathrm{H} \right)$, that is determined from the solution of the following kinetic equation:
\begin{eqnarray} \label{eqBoltz}
\mathbf{w}_{\mathrm{H}} \cdot
\frac{\partial f_{\mathrm{H}}}{\partial \mathbf{r}} +
\frac{\mathbf{F}_\mathrm{r} +
	\mathbf{F}_\mathrm{g}}{m_{\mathrm{H}}} \cdot \frac{\partial
	f_{\mathrm{H}}}{\partial \mathbf{w}_{\mathrm{H}}} = -
\nu_\mathrm{ph}   f_{\mathrm{H}}
\left( \mathbf{r}, \mathbf{w}_{\mathrm{H}} \right) \nonumber \\
-f_{\mathrm{H}} \cdot \int
|\mathbf{w}_{\mathrm{H}}-\mathbf{w}_\mathrm{p}|\sigma_\mathrm{ex}^{\mathrm{HP}}(|\mathbf{w}_{\mathrm{H}}-\mathbf{w}_\mathrm{p}|)
f_\mathrm{p} \left( \mathbf{r}, \mathbf{w}_\mathrm{p} \right)
\mathrm{d}\mathbf{w}_\mathrm{p}  \\
+ f_\mathrm{p} \left( \mathbf{r} , \mathbf{w}_{\mathrm{H}} \right)
\int |\mathbf{w}_{\mathrm{H}}^{\ast } - \mathbf{w}_{\mathrm{H}} |
\sigma_\mathrm{ex}^{\mathrm{HP}}(|\mathbf{w}_{\mathrm{H}}^{\ast } - \mathbf{w}_{\mathrm{H}} |) f_{\mathrm{H}} \left( \mathbf{r} ,
\mathbf{w}_{\mathrm{H}}^{\ast} \right)
\mathrm{d}\mathbf{w}_{\mathrm{H}}^{\ast}. \nonumber
\end{eqnarray}
Here, $\mathbf{F}_\mathrm{r}$ and $\mathbf{F}_\mathrm{g}$ are the forces of solar radiation pressure and the solar gravitation, respectively.
It is assumed in the present calculations  that $|\mathbf{F}_\mathrm{r}|/|\mathbf{F}_\mathrm{g}| = 1.258$. Despite the fact that this parameter (and its temporal variations) is highly important for the distribution of H atoms in the heliosphere inside a sphere of 10-20 AU (e.g., Katushkina et al. 2015), it does not influence on the global structure of the SW/LISM region  discussed in this paper.
$f_\mathrm{p} \left( \mathbf{r},
\mathbf{w}_\mathrm{H} \right)$ is the local Maxwellian
distribution function of protons with gas dynamic values of $\rho
\left( \mathbf{r} \right)$, $\mathbf{V} \left( \mathbf{r} \right)$
and $T \left( \mathbf{r} \right)$.
$\sigma_\mathrm{ex}^\mathrm{HP}(u)$ is the effective charge-exchange cross
section and
$\sigma_\mathrm{ex}^\mathrm{HP}(u)
= \left(2.2835\cdot 10^{-7} - 1.062\cdot 10^{-8} ln(u) \right)^2$ cm$^2$
(Lindsay \& Stebbings 2005), where $u$ is the relative atom-proton velocity in cm s$^{-1}$.
$\nu_{ph} = 1.67\cdot 10^{-7} (R_E/R)^2$  $s^{-1}$,  ($R_E = 1 AU$)
is the photoionization rate.

The charged (plasma) component is described in the context of an ideal MHD approach:
\begin{equation}
\nabla \cdot( \rho \mathbf{V})=q_1 ,
\label{eq-continuity-MHD-section}
\end{equation}
\begin{equation}
\label{eq-momentum-MHD-section}
\nabla\cdot\left[\rho{\mathbf{V}}{\mathbf{V}} + \left(p + \frac{B^2}
{8\pi}\right){\mathbf{I}} - \frac{{\mathbf{B}}{\mathbf{B}}}
{4\pi}\right] = \mathbf{q}_{2},
\end{equation}
\begin{equation}
\label{eq-MF-MHD-section}
\nabla\cdot({\mathbf V}{\mathbf B} - {\mathbf B}
{\mathbf V})=0,
~~\nabla \cdot \mathbf{B} = 0,
\end{equation}
\begin{equation}
\label{eq-energy-MHD-section}
\nabla\cdot\left[\left(E + p + \frac{B^2}{8\pi}\right){\mathbf V} -
\frac{({\mathbf V} \cdot {\mathbf B})}{4\pi}{\mathbf B}\right]
= q_3,
\end{equation}
 where  ${\mathbf B}$ is the magnetic field induction vector,
	${\mathbf a}{\mathbf b}$ is the tensor production of two vectors ${\mathbf a}$ and
	${\mathbf b}$, ${\mathbf I}$ is the unity tensor, '$\cdot$' is the scalar product,
	$E=\frac{\rho V^2}{2} + \frac{p}{\gamma - 1} + \frac{B^2}{8\pi}$, and  $\gamma=5/3$.
	The total plasma density is determined as $\rho= m_p n_p+ m_{He }n_{He}$, where $n_{He}$ denotes the He$^{+}$ number density in the interstellar medium and the He$^{++}$ number density in the SW. To determine the number densities, we solve the continuity equations for the helium ion component in the LISM and for alpha particles in the SW. Then, the proton number density can be obtained as $n_p = (\rho - m_{He }n_{He})/m_p$.  $p$ is the total plasma pressure,  defined as $p = (2n_p+3n_{He^{++}}) k_B T_p$ in the SW and as $p = 2(n_p+n_{He^+}) k_B T_p$ in the LISM. $T_p$ is the plasma temperature and $k_B$ is the Boltzmann constant.
	
The influence of charge exchange with the interstellar H atoms was taken into account on the right-hand side of the MHD equations by the source terms  $q_{1}$,$\mathbf{q}_{2}$, and $q_3$. The sources are calculated as the following integrals of the H-atom velocity distribution function:
\begin{equation}
\label{q1-helium-section} q_{1} = m_{\mathrm{p}} n_{\mathrm{H}}
\cdot \nu_{\mathrm{ph}}  ,
n_{\mathrm{H}} \left( \mathbf{r},t \right) = \int
f_{\mathrm{H}} \left( r, {\mathrm w}_{\mathrm{H}} \right)
\mathrm{d} {\mathrm w}_{\mathrm{H}},
\end{equation}
\begin{equation}
\label{q2-helium-section} \mathbf{q_{2}} = \int m_{\mathrm{p}}
\nu_{\mathrm{ph}}
\mathbf{w}_{\mathrm{H}} f_{\mathrm{H}} \left( \mathbf{r},
\mathbf{w}_{\mathrm{H}} \right)
\mathrm{d}\mathbf{w}{_{\mathrm{H}}} +
\end{equation}
\[
\int \int m_{\mathrm{p}} v_{\mathrm{rel}}
\sigma_\mathrm{ex}^\mathrm{HP} \left( v_{\mathrm{rel}} \right)
\left( \mathbf{w}_{\mathrm{H}} - \mathbf{w} \right) f_{\mathrm{H}}
\left( \mathbf{r}, \mathbf{w}_{\mathrm{H}}, t \right) f_\mathrm{p}
\left( \mathbf{r}, \mathbf{w} \right) \mathrm{d}
\mathbf{w}_{\mathrm{H}} \mathrm{d} \mathbf{w} ,
\]
\begin{equation}
\label{q3-helium-section} q_{3} = \int m_{\mathrm{p}}
\nu_{\mathrm{ph}}  \frac{{\rm
		w}_{\mathrm{H}}^{2}}{2}f_{\mathrm{H}} \left( \mathbf{r} ,
\mathbf{w}_{\mathrm{H}} \right) \mathrm{d}
\mathbf{w}_{\mathrm{H}} +
\end{equation}
\[
\frac{1}{2} \int \int m_{\mathrm{p}} v_{\mathrm{rel}}
\sigma_\mathrm{ex}^{\mathrm{HP}} \left( v_{\mathrm{rel}} \right)
\left( {\rm w}_{\mathrm{H}}^{2} - {\rm w}^{2} \right)
f_{\mathrm{H}} \left( \mathbf{r} , \mathbf{w}_{\mathrm{H}}
\right) f_\mathrm{p} \left( \mathbf{r} , \mathbf{w} \right) \mathrm{d}{\mathbf
	w}_{\mathrm H} \mathrm{d}{\mathbf w}
\]
\[
+n_{\mathrm{H}}  \nu_{\mathrm{ph}} E_{\mathrm{ph}}
\]
Here, $v_{\mathrm{rel}}=\mid \mathbf{w}_{\mathrm{H}}-\mathbf{w}|$
is the relative velocity of an atom and a proton and
$E_{\mathrm{ph}}$ is the mean photo\-ionization energy (4.8 eV).

 Source terms (6)-(8) have been calculated using the  Monte Carlo method (Malama 1991) that was used to solve kinetic equation (1). This method assumes that the velocity distribution function of the proton component is locally Maxwellian, although it allows a generalization to any isotropic distribution function as it was done by Malama et al. (2006). 

Although our model and numerical code allow us to take into account the effects of electron impact ionization that might have an influence on the inner heliosheath plasma flow (e.g., Figure 3 in  Baranov \& Malama 1996) and on interstellar neutral filtration in the SW/LISM interaction region (Izmodenov et al. 1999), we do not take it into account here for the following two reasons. First,  the temperature of the electrons is probably overestimated in the  one fluid plasma approach employed here. A more rigorous multi-component plasma approach is needed (e.g., Malama et al. 2006; Izmodenov et al. 2009). Second,  in the current paper, we will explore the effects of HMF on the plasma heliosheath and, in particular, the effect of plasma density diminishing in the vicinity of the heliopause.  The electron impact (if included) would make this effect less evident.

For the purpose of this paper, we performed model calculations for three different sets of boundary conditions. These sets are chosen in order to clearly explore the effects of the HMF and the dependence of the SW parameters on heliolatitude. Model 1 corresponds to the model where neither the HMF nor the heliolatitudinal dependence of the SW are taken into account. This is simply the model that was published previously (Izmodenov et al. 2005b, 2009) but with updated LISM boundary conditions.
In  model 2, we take into account the HMF, while the SW is still assumed to be spherically symmetric. Model 3 takes into account both the HMF and the heliolatitudinal variations of the SW.

To finish the formulation of the problem, the inner boundary conditions in the unperturbed (by LISM) SW and the outer boundary conditions in the pristine LISM should be formulated.
 It is important to note that the boundary conditions considered in this paper are stationary. Therefore,  this paper focuses on the stationary solutions of the SW/LISM interaction problem. Nevertheless, our numerical method can be easily generalized for time-dependent (periodical) solutions, as was done for the axisymmetrical model by Izmodenov et al. (2005, 2008).

\subsection{Interstellar parameters}

The interstellar H atom and proton number densities are assumed to be n$_{H, LISM}$ = 0.14 cm$^{-3}$ and n$_{p, LISM}$ = 0.04 cm$^{-3}$, respectively. Both densities are somewhat smaller  compared to the values of n$_{H, LISM}$ = 0.18 cm$^{-3}$ and n$_{p, LISM}$ = 0.06 cm$^{-3}$ which were accepted from previous studies (see, e.g., Izmodenov 2009).
The reasons for such a reduction are discussed in  Section 4.

The number density of the interstellar helium ions is chosen as $n_{He+, LISM} = 0.003$ cm$^{-3}$. This corresponds to the standard universal ratio of the total H to He, 	$(n_{p,LISM}+n_{H,LISM})/(n_{He^+,LISM}+n_{He,LISM})=10$ and $n_{He,LISM}=0.015$ cm$^{-3}$ (Gloeckler et al. 2004, Witte 2004). Interstellar helium is taken into account in all three of the models considered in this paper.

The interstellar bulk velocity and temperature are chosen as V$_{LISM}$ = 26.4 km s$^{-1}$, T$_{LISM}$ = 6530 K. This velocity value is consistent with the ISM vector derived from analyses of {\it Ulysses}/GAS data for interstellar helium (Witte 2004, Bzowski et al. 2014, Katushkina et al. 2014, Wood et al. 2015).
The direction ${\bf V}_{LISM}$ is determined by the interstellar helium flow direction from {\it Ulysses}/GAS data. This value is (longitude=-1.02$^{\circ}$, latitude=-5.11$^{\circ}$) in the heliographic inertial coordinate system (HGI 2000).
The velocity vector is in agreement with new results derived from {\it IBEX} data (see, e.g. McComas et al. 2015).
The temperature of T$_{LISM}$ = 6530 K is slightly smaller than the newly adoped temperature of 7500 K (McComas et al., 2015). 


The magnitude of the (IsMF) is assumed to be B$_{LISM}$ = 4.4 $\mu$ Gauss and its direction has $\alpha$=20$^{\circ}$ with the direction of the interstellar bulk flow.  These values were obtained in the parametric study presented in Table 2 of Izmodenov et al. (2009) to obtain asymmetry of the TS agreeable with the distances of the {\it Voyager 1} and {\it Voyager 2} TS crossings.

The chosen values of the velocity and IsMF correspond to the supersonic and subalfvenic LISM. The sonic, alfvenic and fast-magnetosonic Mach numbers
are
M = 2.17, A = 0.631, and M$_{Z+, LISM}$ = 0.628. M$_{Z+, LISM}$ is calculated here along the LISM flow direction.

In the model, it is assumed that the plane which is determined by the vectors ${\bf V}_{LISM}$  and ${\bf B}_{LISM}$  ({\it BV}-plane) coincides with the plane determined by the velocity vectors of the interstellar helium, ${\mathbf V}_{He, LISM}$,  and hydrogen, ${\mathbf V}_{H, LISM}$, inside the heliosphere - the so-called Hydrogen Deflection Plane (see, Lallement et al. 2005, 2010; for model justification see Izmodenov et al., 2005b, 2009). Let us define the positive normal to the plane as
\begin{equation}
{\bf e}_y =  \frac{[{\bf V}_{LISM} \times {\bf B}_{LISM}]}{|[{\bf V}_{LISM} \times {\bf B}_{LISM}]|}
= \frac{[{\bf V}_{He, LISM} \times {\bf V}_{H, LISM}]}{|[{\bf V}_{He, LISM} \times {\bf V}_{H, LISM}]|}.
\end{equation}

Then, for the direction of the helium flow given above and for the hydrogen flow direction of (longitude=-4.66$^{\circ}$, latitude=-8.37$^{\circ}$) in HGI 2000 (Lallement et al. 2010), the angle between ${\bf e}_y$ and the north solar pole direction is 137.5$^{\circ}$.
The plane and the angle $\alpha$ within this plane determine the direction of IsMF. In HGI 2000, this direction is (longitude=-16.48$^{\circ}$, latitude=-18.22$^{\circ}$).

\subsection{SW Parameters at Earth's Orbit}

The inner boundary for our model is located at 1 AU. For the models 1 and 2, the SW is assumed to be spherically symmetric. Its number density and velocity are assumed to be n$_{p, E}$ = 5.94 cm$^{-3}$ and V$_{R, E}$ = 432.4 km s$^{-1}$.  These values are obtained by time-averaging the OMNI SW data. We also assume that the number density of the alpha particles $He^{++}$ is 3.5\% of the proton number density.
	
The SW is hypersonic, and we assume that the Mach number is 6.44, which corresponds to a temperature of T$_{E}$ = 188500 K. Our model results are not sensitive to this parameter.

In model 3, we take into account the heliolatitudinal variations of the SW density and speed at 1 AU obtained by time-averaging different experimental data sets.
 Namely, three sets of experimental data are used.
 \begin{enumerate}
   \item In the ecliptic plane, we use data (SW density and speed) from the OMNI 2 database. The OMNI 2 data set contains hourly resolution SW magnetic field and plasma data from many spacecraft in geocentric orbit and in orbit about the L1 Lagrange point.
   \item Heliolatitudinal variations of the SW speed are taken from analysis of the interplanetary scintillation (IPS) data. We
    use the results of Sok\'{o}\l\l ~et al. (2013) where one-year average latitudinal profiles of the SW speed with a
resolution of 10$^{\circ}$ are obtained. Data are available from 1990 to 2011.
   \item Heliolatitudinal variations of the SW mass flux are derived from analysis of {\it SOHO}/SWAN full-sky maps of the backscattered Lyman-alpha
   intensities (Qu\'{e}merais et al. 2006, Lallement et al. 2010, Katushkina et al. 2013). An inversion procedure (see Qu\'{e}merais et al. 2006 for details)
   allows us to obtain the SW mass flux as a function of time and heliolatitude with a temporal resolution of approximately 1 day and an angular resolution
   of 10$^{\circ}$. Data are available from 1996 to 2011.
 \end{enumerate}

\begin{figure}[t!]
\includegraphics[width=0.4\textwidth,clip=]{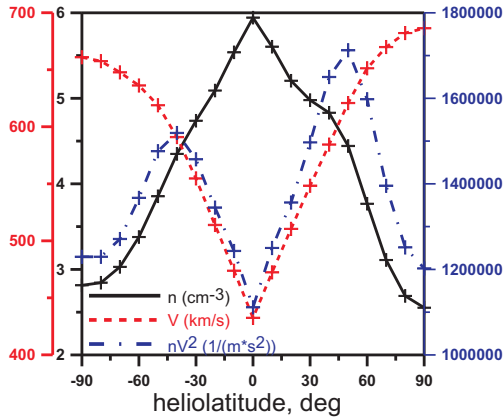}
\caption{Solar wind proton number density and velocity at 1 AU (plot (A)) and dynamic pressure (plot (B)) as functions of the heliolatitude.}\label{sw_1AU}
\end{figure}

From these data, we calculate the mass and momentum fluxes of the SW as functions of time (from 1996 to 2011) and heliolatitude. After that, we average these fluxes over time and then extract the heliolatitudinal variations of the SW density and speed from the averaged fluxes (see results in Figure \ref{sw_1AU}). In the model, it is assumed that the thermal pressure of the SW at 1 AU is constant at all heliolatitudes. Therefore, corresponding heliolatitudinal variations of the SW temperature are applied (variations are about $\pm 20~\%$).  Since the SW is hypersonic, the assumption on thermal pressure at 1 AU does not influence the results.

For the HMF, the Parker spiral solution has been assumed at 1 AU:
\begin{equation}
\label{spiral}
B_R=\pm B_E \left( \frac{R_E}{R}\right)^2,
~B_{\theta}=0,
~B_\varphi=-\beta_s B_R\left(\frac{R}{R_E}\right)sin\theta.
\end{equation}
Here, $(R,\theta,\varphi)$ are the spherical coordinates connected with the 
solar equatorial plane: $\theta$ is the solar latitude, counted from
north solar pole ($0^{\circ}$) to south ($180^{\circ}$).
For $R=R_E=1AU$, we assume $B_E = 37.5 \mu G$, and for solar rotation
parameter $\beta_s = \Omega R_E/V \approx 1$ for the SW velocity $V = 432$ km s$^{-1}$. $\Omega$ is the Sun angular velocity correponding to
the solar rotation period of 25 days.
$\varphi$ is the solar longitude.

The sign of the $B_R$ component should be chosen depending on the location of the heliospheric current sheet that is changing with time.
However, this is not important for our study because the terms which are responsible for the influence of
the magnetic field in the ideal MHD equations do not depend on the orientation of the magnetic field.
Also, in the frame of the ideal MHD equation, we are not interested in the structure of the heliospheric current sheet assuming that it is a discontinuity where the magnetic field changes its orientation.
This assumption has been confirmed so far by {\it Voyager} magnetic field measurements in the SW and in the inner heliosheath (e.g., Burlaga \& Ness 2012) that demonstrate a quick change of the magnetic field orientation rather than a thick (comparing to the size of the heliosheath) transition region.

In addition, the reconnection of the HMF and IsMF at the heliopause is not possible in the frame of the ideal MHD approach. We consider the heliopause as a tangential discontinuity where ${B}_{n} = 0$ on both sides of the heliopause.

Overall, the solution to the considered problem does not depend on the orientations of the inner and outer magnetic fields along their magnetic field lines.
Therefore, to perform numerical calculations, we can assume unipolar HMF at Earth's orbit
$B_R=- B_E \left( \frac{R_E}{R}\right)^2$. The actual polarity of the HMF can be calculated after the self-consistent solution obtained in a similar manner as in  Barsky et al. (1999).

\subsection{Coordinate System}

Henceforward in this paper, we will not use the solar equatorial coordinates. Instead, we use the system of coordinates connected with the
interstellar flow and magnetic field vectors, ${\mathbf V}_{LISM}$ and ${\mathbf B}_{LISM}$.
Axis  $Z$ is directed toward the interstellar flow, i.e., opposite to  ${\mathbf V}_{LISM}$.
Axis $X$ is in the plane containing the ${\mathbf V}_{LISM}$ and ${\mathbf B}_{LISM}$ vectors ({\it BV}-plane) and perpendicular to the Z-axis.
The direction of the X-axis is chosen such that the projection of ${\mathbf B}_{LISM}$ to the X-axis is negative.
Axis $Y$ completes the right-handed system of the coordinates.
The direction of the solar North pole in this coordinate system is  (0.6696, -0.7373, 0.089).

\subsection{Details of the Numerical Scheme and Computational Grid}

In order to obtain a self-consistent steady-state solution of the system (1)-(8) we use the global iteration method (Baranov \& Malama 1993). Non-stationary versions of MHD equations (2)-(5) are solved using a 3D modification of the finite-volume Godunov-type scheme that employs a Harten-Lax-van Leer Discontinuity (HLLD) MHD Riemann solver (see, for example, Miyoshi \& Kusano 2005). A Chakravarthy-Osher TVD limiter with the  possibility to choose compression parameters
(Chakravarthy \& Osher 1985) is used to increase the resolution properties of the
first-order accuracy scheme.

The solver has been adopted to a 3D moving grid with the possibility to fit discontinuities.
We use the soft fitting technique  proposed by Godunov et al. (1979) and
 applied for two-dimensional (2D) MHD flows by Myasnikov (1997).
The chosen numerical scheme allows us to fit all of the major discontinuities - the heliopause, the TS and the bow shock (BS), when the latter exists. All of the other discontinuities (if any) can be captured by the scheme.

 A steady-state solution of Equations (2)-(5) is obtained using the time-relaxation method. In other words, we solve non-stationary analogs of Equations (2)-(5) with stationary boundary conditions for long periods of time when the temporal derivaties of all of the parameters become negligiblely small and the flow pattern reaches steady state in the computational domain. It also has been checked that the obtained steady-state solution remains in a  steady state over very long periods of time, and  so possible instabilities do not destroy the numerical solution.

To satisfy the condition of  $\nabla\cdot \mathbf{B}=0$ in our numerical solution of the stationary problem,
we follow the well-known procedure suggested by Powell et al. (1999), which consists of adding to the right parts of the 
non-stationary versions of the MHD Equations (2)-(5) those terms which are
proportional to $\nabla\cdot \mathbf{B}$. In the momentum equation, the term
 $[-\mathbf{B}(\nabla\cdot \mathbf{B})/4\pi]$ has been added. In the
 energy equation, the term $[-(\mathbf{V}\cdot\mathbf{B})(\nabla\cdot \mathbf{B})/4\pi]$ has been added.
In the magnetic field induction equation, we have added the term $[-\mathbf{V}(\nabla\cdot \mathbf{B})]$.

In the calculations, we use a specific non-regular moving grid that allows us to perform exact fitting of the TS and heliopause. This grid allows us to decrease the sizes of the cells (e.g., increase the number of cells) in the vicinity of the discontinuities.

The example of a typical numerical grid  in the (ZX) and (XY) planes  is shown in Figure 2.
 The convergence of the numerical solution has been verified  by  changing the grid resolution up to two times in all directions and
 by changing cell sizes specifically in the vicinities of the discontinuities.
Also, the  numerical solution has been tested on the extended grids where the computational domain was increased up to two times.
In particular, the tail region for model 3  varied in these calculations from 1000 AU up to 5500 AU.

\begin{figure}[t!]
	\centerline{\hspace{0.3cm}\includegraphics[width=\textwidth,clip=]{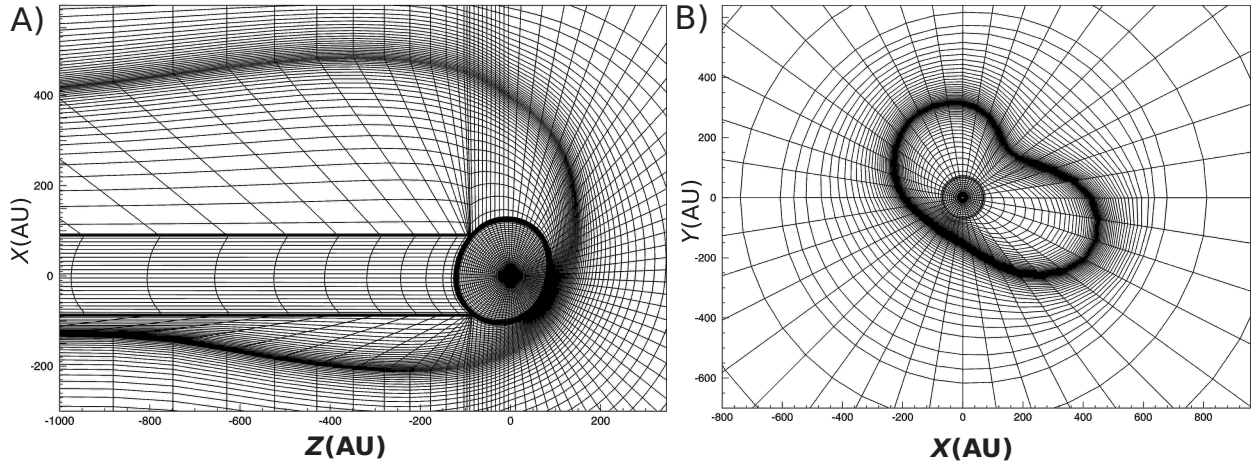}}
	\caption[1] {Demonstration of the computational grid in the ZX plane (panel (A)) and in the plane parallel to the XY plane at z=-500 AU (panel (B)).}\label{fig-grid}
\end{figure}

 Kinetic Equation (1) for the H-atom component has been solved using a Monte Carlo method with spatial splitting of the particle trajectories (Malama, 1991). Monte Carlo calculations have been performed  using the same computational grid as described above for the plasma component.
Since the TS and heliopause are the grid surfaces, we can define different populations of H atoms through their dependence on their place of origin (see, e.g., Malama 1991; Baranov et al. 1998). Optimization of the critical weights in the Monte Carlo method has been performed separately for the each of the four populations (Malama 1990). This procedure  increased the  efficiency of the method.

\section{Results}

\begin{figure}
\centerline{\hspace{0.3cm}\includegraphics[width=\textwidth,clip=]{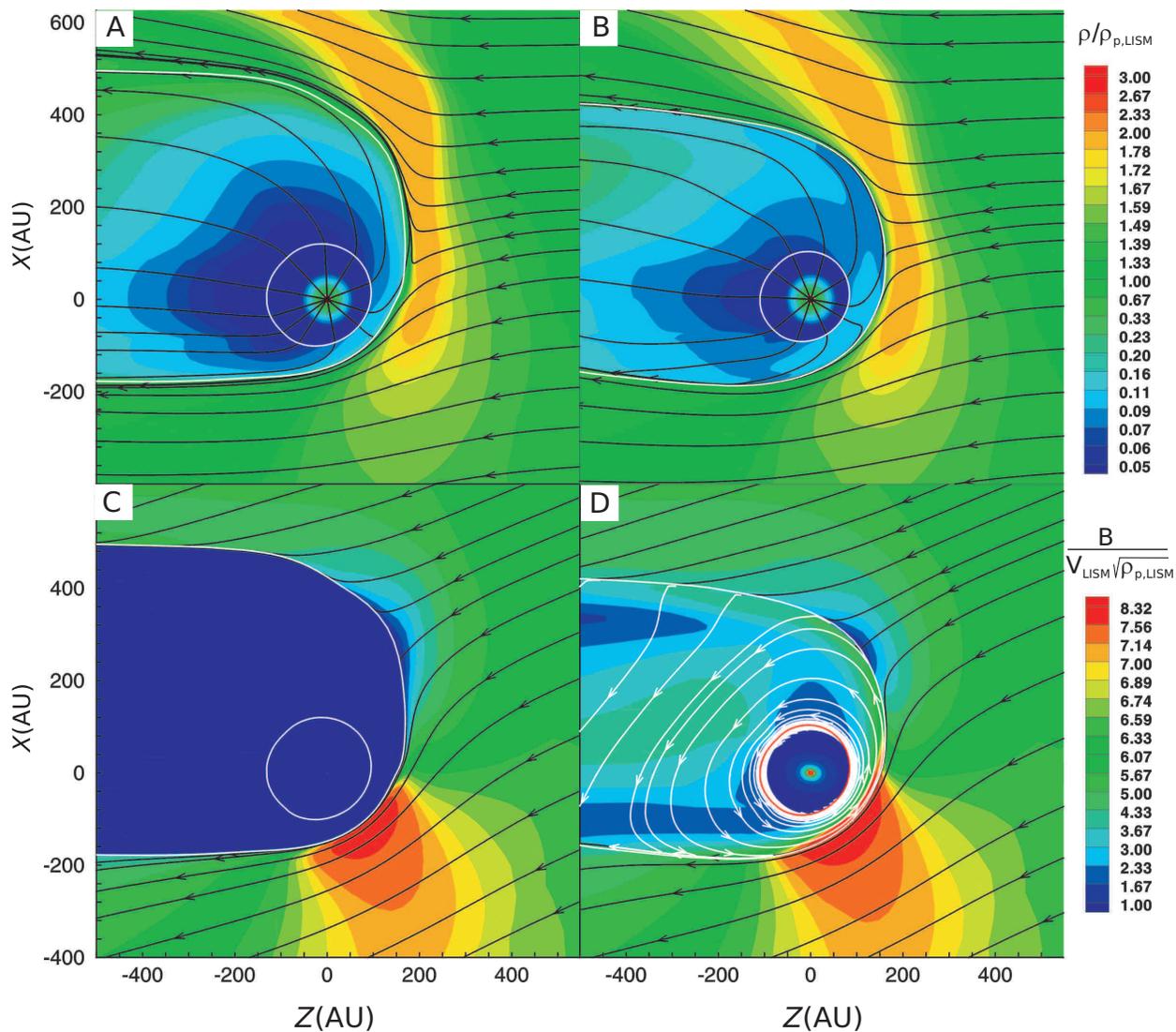}}
\caption[1]{Panels (A) and (B): plasma streamlines and isolines of the plasma density normalized to the proton density in LISM. Panels (C) and (D): magnetic field lines and isolines of the magnetic field magnitude in dimensionless units. Left panels ((A) and (C)) correspond to Model 1 without HMF, right panels ((B) and (D)) correspond to Model 2 with HMF. All of the panels are made in the ZX plane determined by the interstellar velocity and magnetic field vectors.}\label{fig-interface}
\end{figure}

\begin{figure}
	\includegraphics[width=\textwidth,clip=]{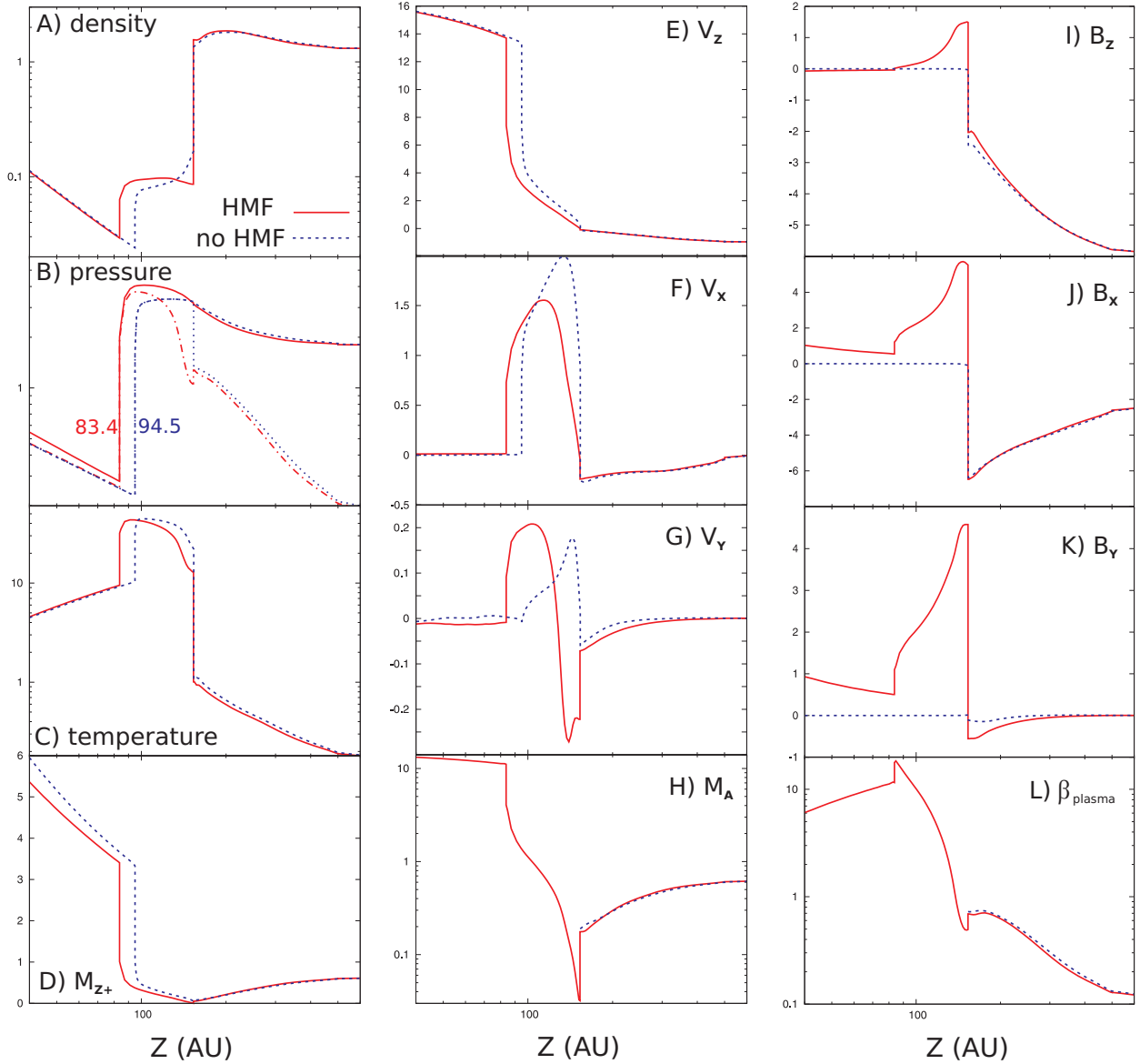}
	\caption{Plasma and magnetic field parameters in the upwind direction (Z-axis) for Model 1 (dashed blue curves) and Model 2 (solid red curves): (A)  plasma density, (B) thermal (dash-dot curves) and total pressures, (C) plasma temperature, (D) fast magnetosonic Mach number calculated for the direction along the Z-axis; (E),(F),(D)  V$_z$, V$_x$, V$_y$ components of the plasma velocity; (H)  Alfvenic number; (I),(J),(K)  B$_z$, B$_x$, B$_y$ components of the magnetic field; (L)  plasma beta. \label{1d-upwind}}
\end{figure}

\begin{figure}
	\includegraphics[width=\textwidth,clip=]{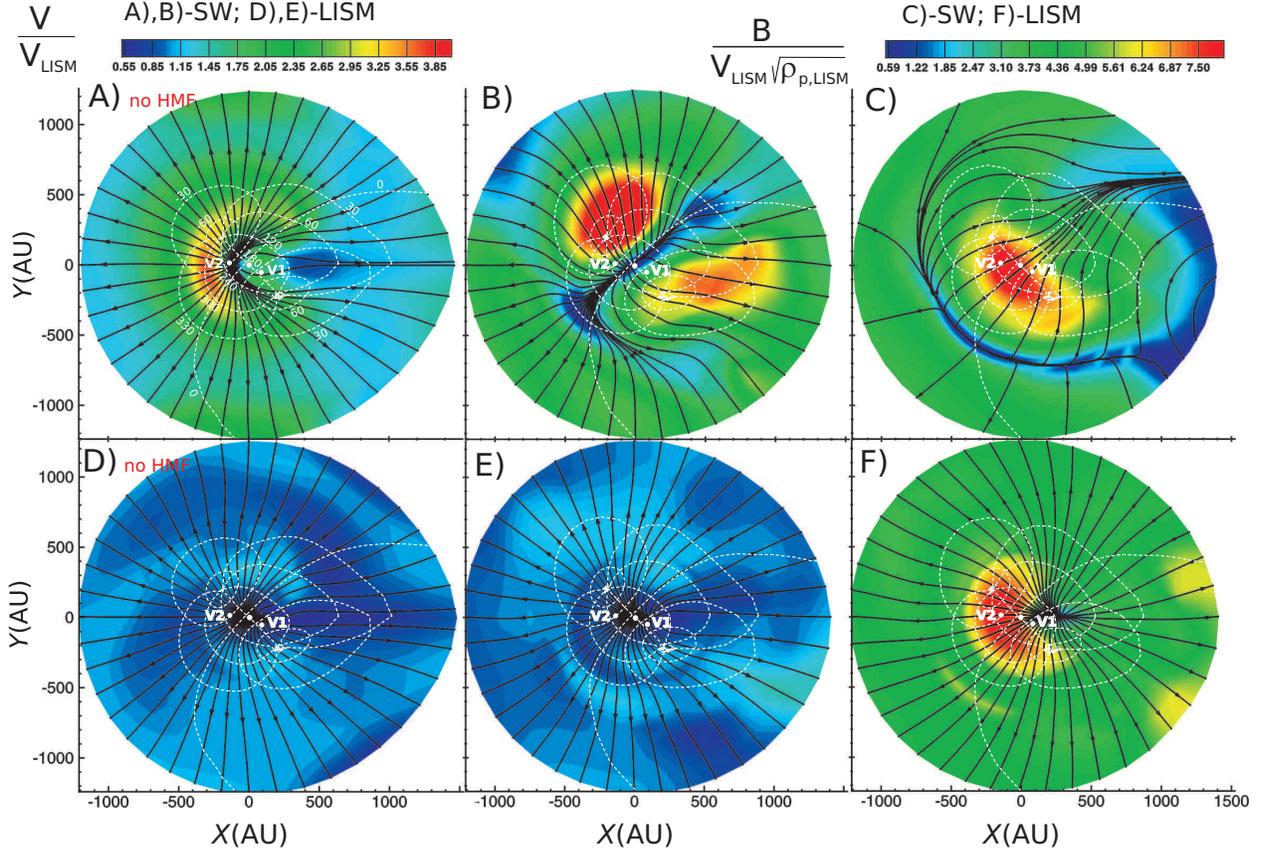}
	\caption{ Solar wind and interstellar velocities and magnetic fields along the heliopause. A description of the projection plane is given in the text.  Panels (A) and (B): the solar wind velocity streamlines and isolines of the solar wind velocity magnitude for Models 1 and 2, respectively.
		Panel (C): HMF lines and isolines of the HMF magnitude for Model 2.
		Panels (D) and (E):  the interstellar plasma velocity streamlines and isolines of the solar wind velocity magnitude for Models 1 and 2, respectively. Panel (F):  the IsMF field lines and isolines of the HMF magnitude for Model 2.
		 White dashed lines are the lines of latitude and longitude in the HGI 2000 coordinate system. Lines of latitude marked as $\pm$30 and $\pm$60 are centered around the poles, respectively. Line "0" is the projection of the solar equatorial plane. Lines of longutude pass through the poles. They are marked by their longitudes of 30, 60, 120, 180, 240, 300, and  330, respectively. The upwind, {\it Voyager 1} and {\it Voyager 2} directions are marked as Z, V1, and V2.
  \label{fig3}}
\end{figure}

\begin{figure}
	\includegraphics[width=\textwidth,clip=]{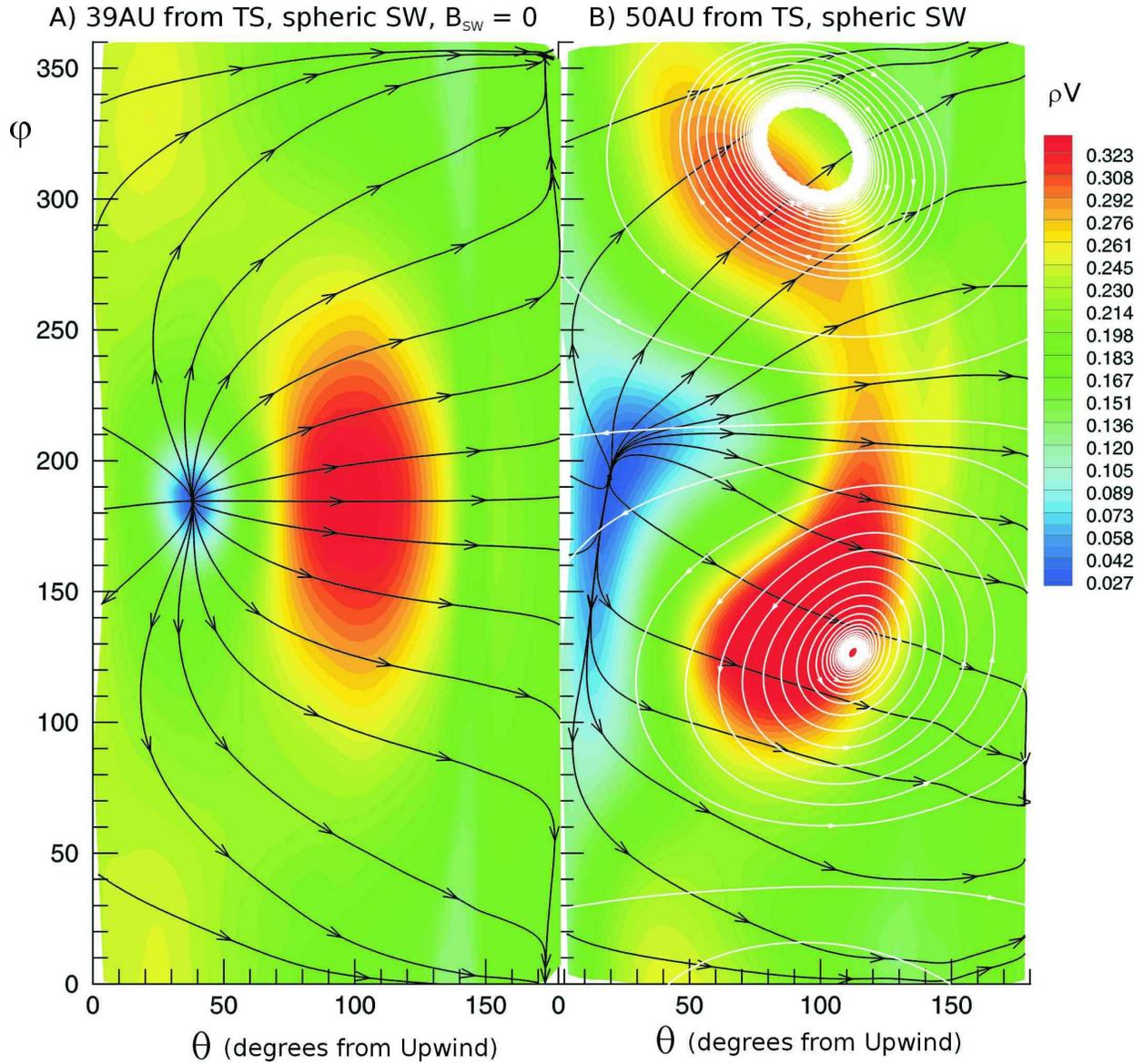}
	\caption{ Solar wind mass flux ($\rho V$)  projected on a closed surface located in the inner heliosheath at equal distances from the heliopause:  (A) Model 1 with no HMF, (B) Model 2. The angles are $\theta$ and $\phi$  Black curves are the projections of the streamlines into the surface. White curves are the projections of the heliosheric magnetic field lines.  }  \label{fig6}
\end{figure}

\begin{figure}
	\includegraphics[width=\textwidth,clip=]{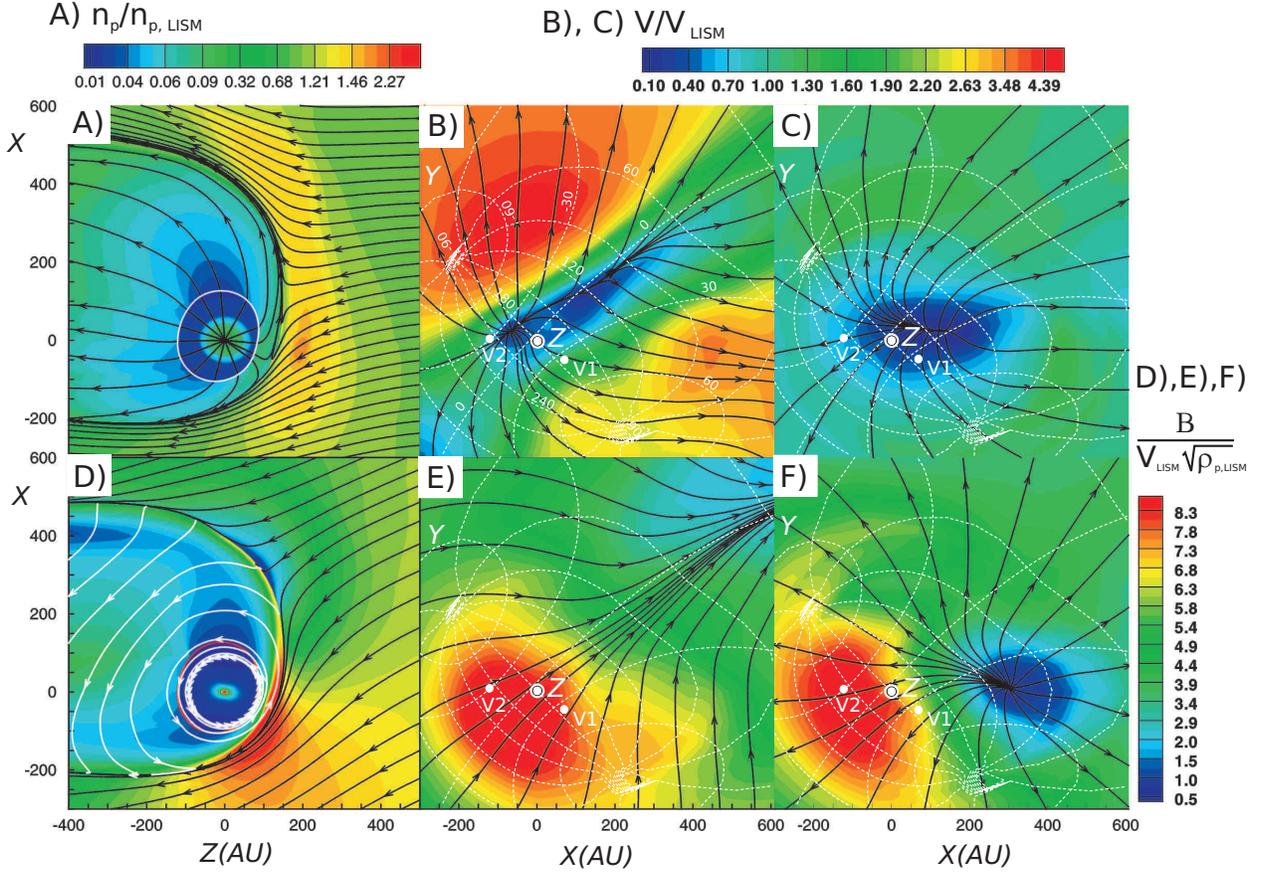}
	\caption{ Plasma and magnetic field distributions obtained in the context of Model 3, in which both HMF and heliolatitudinal variations of the solar wind have been taken into account. Panels (A) and (D) show the plasma streamlines and density isolines (panel (A)) and the magnetic field lines and HMF isolines (panel (D)) in the ZX plane. Panels (B),(C),(E), and (F) present the plasma velocity and magnetic field distributions along the heliopause  on the heliospheric (panels (B) and (E)) and  interstellar (panels (C) and (F)) sides of the heliopause. The projection plane is the same as in Figure 5.  White dashed lines are the lines of latitude and longitude in the HGI 2000 coordinate system (see, Fig.5).   \label{fig7}}
\end{figure}

\begin{figure}
	\includegraphics[width=0.8\textwidth,clip=]{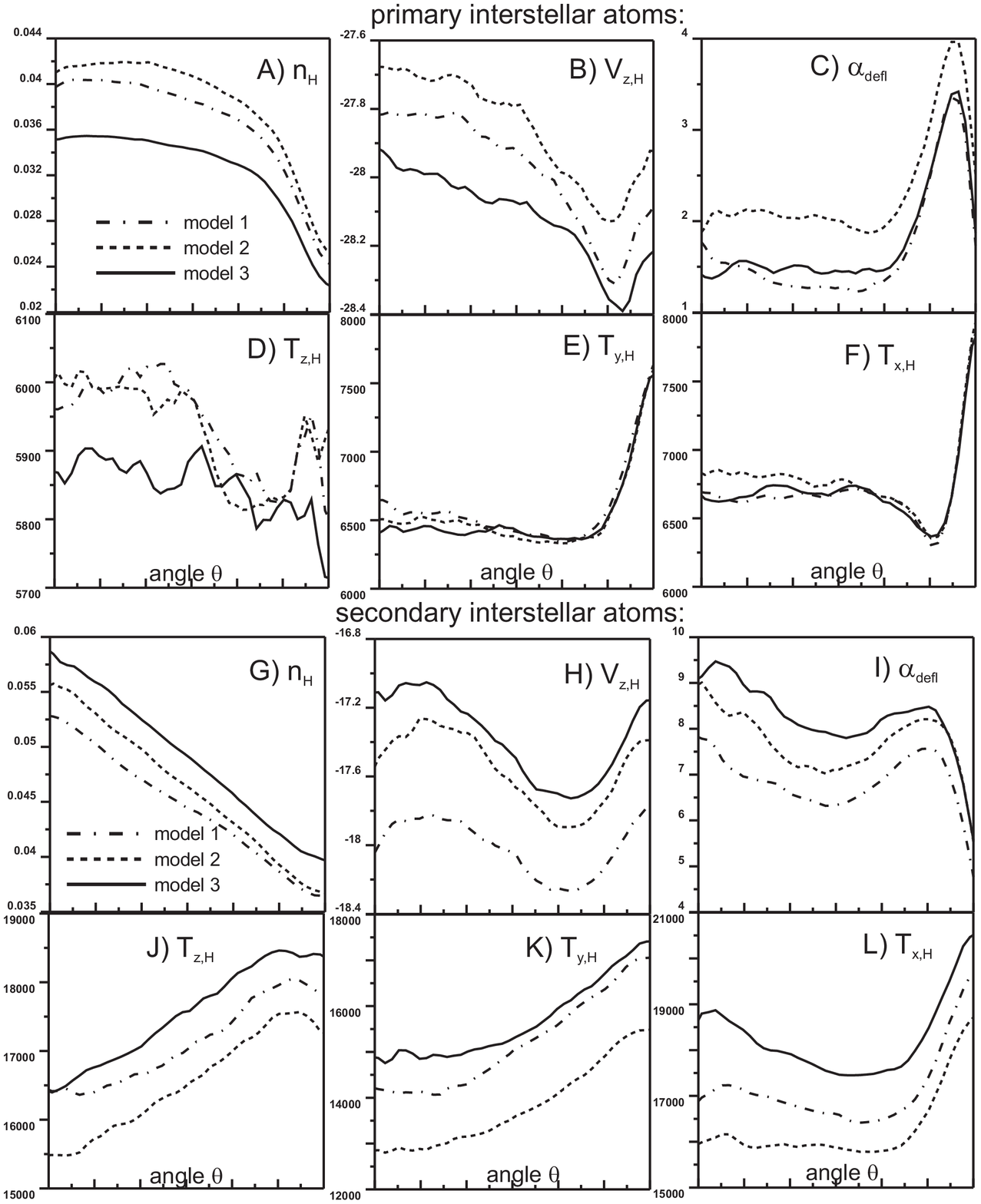}
	\caption{ Parameters of the primary (A)-(F) and secondary (G)-(L) interstellar atom populations at 90 AU as a function of the angle $\theta$ measured from the upwind direction in the ZX plane. Panels (A) and (G): number density in cm$^{-3}$. Panels (B) and (H): Z component of velocity in km s$^{-1}$. Panels (D)-(F), and (J)-(L): components of kinetic temperatures. Plots (C) and (I) show the deflection angle $\alpha_{defl}$ that characterizes the deflection of the H-atom velocity from the unperturbed (by the interaction with the Sun) interstellar velocity vector.
	 \label{fig7}}
\end{figure}

First, we will explore the effects of the HMF.
Figure 3 shows the results of numerical simulations for plasma in the ZX plane (see Section 2.3 for definition).
Panels (A) and (C) present the results obtained in the frame of Model 1 (without HMF), while  panels (C) and (D) represent Model 2 (with HMF).
The panels on the top of the figure (panels (A) and (B)) show the plasma streamlines and the isolines of the plasma number density, while the panels at the bottom (panels (C) and (D)) show the magnetic field lines and isolines of HMF's magnitude.

Visual comparison of the results shows that the HMF does not qualitatively change the global structure of the interaction region. The geometrical patterns and distributions of the plasma density and magnetic field look quite similar, especially outside of the heliopause.

However, it can clearly be seen that the heliospheric TS is about $\sim$10 AU closer to the Sun in Model 2  compared to Model 1.
This effect is quantitatively seen in Figure 4, which shows one-dimensional (1D) distributions of the plasma and magnetic field in the direction of the Z-axis, i.e. toward the vector of the interstellar flow or in the upwind direction.
In the upwind direction (and in the entire region close to the SW stagnation point that is shifted south of X=0), the heliosphere is located at the same distances for Models 1 and 2.
However, the further one moves from the upwind region, the larger the difference between the HP distances in Models 1 and 2. In the crosswind (i.e., in the direction of positive X) the HP is about 100 AU further in Model 1 than in Model 2.

This effect could be explained by a plasma depletion layer around the heliopause. The effect of plasma depletion in the vicinity of contact discontinuities is known for the magnetospheres (e.g., Zwan \& Wolf 1976). The effect is connected with an increase in the perpendicular component of the magnetic field when plasma approaches the contact boundary (Cranfill-Axford effect for the heliosphere). The gradient of the magnetic field forces plasma to decelerate and  flow around of the magnetic ``wall" obstacle. This effect can clearly be seen in Figure 4 where the 1D distributions of the plasma and magnetic field parameters are shown for the upwind direction. All three components of the  HMF increase toward the heliopause (plots (I),(J), and (K) in Figure 4). The plasma velocity components (plots (E),(F), and (G)) and density (plot (A)) decrease from the middle of the inner heliosheath to the heliopause. In the same region, both the Alfvenic number and plasma beta become less than 1, i.e., the flow is determined by the magnetic field.
The effect of SW plasma density depletion around the heliopause can clearly be seen from the comparison of plasma density isolines in Models 1 and 2 (compare panels (A) and (B) in Figure 3).

Magnetic field pressure replaces the plasma pressure in such a way that the total pressure is conserved (plot (B) in Figure 4). This conclusion is valid for quite a wide region around the heliopause. However, because less plasma flows along the heliopause from the stagnation region to the flanks, some deficit of pressure appears at the flanks and the heliopause there moves toward the Sun.
This explains why  the heliopause in Model 2 has a slightly different (from Model 1) shape  and is closer to the Sun in the crosswind and tail.

It is interesting to note that a sharp and strong jump of density at the heliopause exists in Model 2. This somewhat contradicts the Belov and Ruderman (2010) result for the smooth transition at the heliopause due to the  charge-exchange effect.
We explain this  by the fact that the effect does not work in 3D geometry since the stagnation points are in different places inside and outside the heliopause.

We also note that we discuss the plasma depletion from the inner side of the heliopause. A similar effect of plasma depletion should be pronounced from the outside of the heliopause, as discussed in Fuselier \& Cairns (2013).
We do not see the effect in the model because the interstellar H atoms strongly influence (due to charge exchange) the interstellar plasma flow around the heliopause. However, some small decrease of the plasma density that is observed in Figure 4(A) at the outer side of the heliopause could be associated with the plasma depletion.

The effect of the magnetic increase (or magnetic wall) around the heliopause (plot (D) in Figure 3) and the plasma depletion leads to quite significant displacement of the heliospheric TS toward the Sun. This can be  seen in Figure 4(B). In Model 2, the TS is $\sim$11 AU closer to the Sun  compared with Model 1.
Such a displacement of the TS is probably the major effect of the HMF on the global structure of the heliosphere.

Despite the fact that the geometrical pattern of the SW/LISM interaction region does not change dramatically, the HMF significantly influences the SW plasma flow. To illustrate this,  we present the distribution of the plasma velocity and magnetic field along the heliopause (Figure 5).
In this figure, the heliopause is projected onto a (projection) plane as follows.
The origin of the coordinate system of this plane coincides with the intersection of the heliopause with the Z-axis. Polar coordinates ($r$, $\varphi$) in this plane are defined in such a way that the polar angle  $\varphi$ coincides with the polar angle in the (XY) plane.
The radial distance $r$ coincides with the length of the curve that is obtained at the intersection of the heliopause and the plane containing the Z-axis and determined by constant $\varphi$ angle in the (XY) plane.

Note that the ({\it BV}) plane (i.e., ZX plane) corresponds to the horizontal line $Y=0$ in the projection plane. For Model 1, the plasma flow pattern is symmetric around this line because the (ZX) plane is the plane of the symmetry in this model. This is not the case for Model 2. The main symmetry is observed around line $\varphi$ = 42.4$^\circ$.  This line is defined by the angle between the solar equatorial plane and the ({\it BV}) plane. Therefore, the SW plasma flow pattern in the heliosheath (and, specifically, at the heliopause) is determined by the direction of the solar rotation axis rather than by the directions of the interstellar flow and magnetic field vectors.
It can be seen (panel (B) in Figure 5) that the SW velocity is smaller in the equatorial plane and the streamlines directed toward the poles.
Some asymmetries with respect to the plane of  $\varphi$ = 42.4$^\circ$ are observed in the figure. This is due to the asymmetric shape of the heliopause distorted by the IsMF. However, these "interstellar" induced features are much less pronounced than the "Sun-connected" symmetry of the flow.

The same types of symmetries/asymmetries are also observed in the distribution of the HMF along the heliopause (panel (C) of Figure 5).

As opposed to the inner side of the heliopause, in the interstellar medium, the plasma and magnetic field distributions look very similar in Models 1 and 2 (panels (D)-(F) in Figure 5). Small differences in panels (D) and (E) are caused by the differences in the heliopause shape in Models 1 and 2.

 Before we move forward to discuss the results obtained in the context of Model 3, it is worthwhile to mention the very recent publication by Opher et al. (2015), where it was pointed out that the HMF tension-force in the inner heliosheath leads to the formation of two SW flux jets directed  north and south. Furthermore, this tension force changes the topology of the heliopause, making it a more tube-like structure (see, also,~ {\bf axisymmetric model by} Drake et al. 2015) as compared to our sheet-like structure. The model by Opher et al. (2015) assumes a unipolar magnetic field, as in our Model 2. Therefore, the effects of magnetic tension in the heliosheath can be seen in our Model 2. In order to explore these effects, we plot (Figure 6) the solar flux ($\rho V$) in a projection on a closed surface that it located in the inner heliosheath. The surface has been constructed by adding a fixed distance to the TS distance in any given direction.  The distance has been chosen as 39 AU for Model 1 and 50 AU for Model 2. The difference in the distances chosen for Models 1 and 2 is connected with the different widths of the inner heliosheath in the models.
 Figure 6 presents the solar flux plotted for Model 1 and Model 2.
Any point on the surface is characterized by two spherical angles $\theta$ and $\phi$. The angle $\theta$ is counted from the Z-axis (i.e., from the upwind direction), while the angle $\phi$ is counted from the X-axis in the XY plane.

Panel (1) in Figure 6 demonstrates the SW mass flux in the case of Model 1 when HMF is not taken into account. It is seen that the mass flux reaches its maximum  for $\theta$ $\sim$100$^{o}$-110$^{o}$ and $\phi$$\sim$180$^{o}$.
This nearly crosswind direction corresponds to the direction through which all of the solar mass flux from the upwind hemisphere passes into the tail. The maxima at $\phi$$\sim$180$^{o}$ corresponds to that part of the ({\it BV}) plane where the distance between the TS and HP is minimal.  Conversely, panel (2) demonstrates that there are two mass-flux maxima (i.e. jets) in the north and south directions. These jets are clear results of the magnetic tension. Therefore, in this respect, our model results is in a qualitative agreement with the conclusion of Opher et al. (2015).

However, the  conclusion on the tube-like shape of the heliopause made by Opher et al. (2015) is not supported by our calculations.
The exact reasons for this will be explored in future studies. Here, we speculate on two possibilities.
 The first is connected with effects of charge exchange of solar protons and interstellar H atoms. In the {\bf axisymmetric} model of Drake et al. (2015), {\bf where the spherically symmetric SW flows into the unmagnetized LISM resting with respect to the Sun, the axis of symmetry is the north-south pole axis. Due to the symmetry in any plane containing the axis of symmetry, there is a stagnation point at equator. Therefore, in 3D space, } the stagnation line (i.e., the line of stagnation points) at the heliopause is a circle in the solar-equatorial plane. Since  {\bf the tube-like structure remains} in the case of Opher et al. (2015) when the relative Sun/LISM motion is taken into account, the stagnation line still exists, although it is deformed and not necessarily located in the equatorial plane.
 The interstellar H atoms penetrate into the stagnation line vicinity and begin charge exchange with the protons. As a result of the charge exchange, momentum is transferred from the H atom components into the plasma components. The momentum transfer results in a diminishing the heliocentric distance to the heliopause in the upwind side of the heliosphere. The effect is very well known for the traditional sheet-like heliosphere (see, e.g., Figure 2 in Izmodenov 2000). In the tube-like heliosphere, the same momentum-transfer effect should result in the increase of the heliocentric distance to the heliopause in the downwind direction. The heliopause, in principle, could move further and further downwind,  and so it can be treated as a sheet-like structure  in the considered computational domain. {\bf However, while the effect of charge exchange is, in principle, taken into account in the multi-fluid approach,}  this effect should be verified by kinetic-MHD calculations.
{\bf As  has been shown by Alexashov and Izmodenov (2005), the  multi-fluid and kinetic approaches may have qualitatively different results.}

  Another possible reason Model 2 does not have the tube-like shape of the heliopause could be due to the fact that our computational grid explicitly assumes the sheet-like structure of the heliopause. However, if the formation of the two streams would physically result in the tube-like shape geometry of the heliopause, then at some specific locations (where the SW mass flux becomes negligible) the different parts of our "sheet-like" heliopause should approach each other. Then, our computational grid should collapse and this would be an indication of the tube-like topology of the heliopause.
  However, the grid collapse is not observed in our calculations. Therefore,  we currently favor the first possibility, but the question of the  topology of the heliopause is generally very interesting and should be explored in detail in the future.

Figure 7 presents the results obtained in the framework of Model 3 where the latitudinal variations of the SW parameters have been taken into account in addition to the HMF. It can be seen that both the global structure of the heliosphere and the plasma and magnetic field distributions are qualitatively the same as for Models 1 and 2.
However, the comparison of panel (A) in Figure 7 with panel (B) in Figure 3 shows that the heliopause is more blunt in Model 3. The heliopause is closer (by about~ $\sim$10-15 AU) to the Sun in the upwind part of the heliosphere and further from the Sun in the crosswind.
We associate this with the non-monotonic heliolatitudinal distribution of the SW dynamic pressure, with 30-50 \% increase at the middle latitudes and further decrease toward the poles (see Figure 1).
Such  behavior of the dynamic pressure changes the plasma flow beyond the TS in such a way that more plasma evacuates to the flanks which makes the pressure  smaller in the upwind part and  larger further from the upwind part.

Following the heliopause shape, the TS is also more elongated toward the solar poles. Since it is quite difficult to see this effect in the large scale of Figure 7(A), we present a table of the TS and heliopause positions in the upwind direction, and in the {\it Voyager 1} and {\it Voyager 2} directions. It can be seen that in the upwind direction the TS is only slightly ($\sim$ 1 AU) further away in Model 3  compared with Model 2.
The effect is much larger in the directions of {\it Voyager 1} and {\it 2}. The differences are 9 and 6 AU, respectively.

Panels (B), (C), (E), and (F) in Figure 7 show the projections of  plasma velocities and magnetic fields along the heliopause. Panels (B) and (E) show the distribution inside the heliopause, panels (C) and (F) show it outside the heliopause. All of the main features are essentially the same as in Model 2. Inside the heliosphere, the flow is oriented in accordance with the solar equatorial plane. Although north-south asymmetry can clearly be seen in Figure 7(B).  From the interstellar side of the heliopause, the flow and the magnetic field are oriented along the Y=0 axis which is the projection of the ({\it BV}) plane. Some asymmetries are due to non-symmetric shape of the heliopause.

Since the shape of the heliopause has been changed in Model 3 and the heliopause can be considered as an obstacle to the interstellar flow, the interstellar plasma streamlines are somewhat different in Models 2 and 3. This is seen from the comparison of Figures 7(A) and 3(B). The interstellar atoms are coupled with the interstellar plasma by charge exchange. Therefore, the change in the plasma streamlines could influence the distribution of interstellar H atoms inside the heliosphere. 
Despite the fact that these effects are quite important, they are not visible in a global-scale picture. Qualitatively, the distributions of the interstellar H atoms in the ({\it BV}) plane for Models 2 and 3 look very similar to Model 1. The distributions can be found in Izmodenov \& Baranov (2006, Figure 4.7(B)) or Izmodenov et al. (2009, Figure 5(E)). To explore the difference in H atom distribution quantitatively, we present (Figure 8) the zero, first, and second moments of the velocity distribution function of H atoms at 90 AU as functions of the angle $\theta$ that is counted in the ZX plane from the positive direction of Z (i.e., from the upwind direction).
It is worthwhile to note here that it is convenient to separate the interstellar atoms by primary and secondary populations. The primary population is the original interstellar atoms that penetrate into the heliosphere without charge exchange with protons. The secondary population is the atoms created in the disturbed LISM around the heliopause. Historically, in the case of the supersonic LISM, the atom is defined as a secondary atom if it is created by charge exchange in the region between the BS and the heliopause. Such a definition does not work for models with no BS, such as the models considered in this paper. In this case, we identify a newly originated H atom as the secondary atom if the interstellar plasma is sufficiently disturbed at the place of the atom's origin. More precisely, under sufficiently disturbed conditions, we assume that the temperature of the plasma is more than~ $\sim$10000 K.

Although the total (primary + secondary) number density of the interstellar atoms remains the same for all of the considered models, the proportions of the primary and secondary populations are slightly different (Figure 8(A) and 8(G)). The changes in the interstellar plasma flow around the heliopause (due to different shape of the heliopause) lead to small (less than 1 km s$^{-1}$) changes in the velocities of the primary and secondary components (Figure 8(B) and 8(H)). The kinetic temperature components (which are defined as the second moments of the velocity distribution) for the secondary H atoms are changed within 1000-2000 K. This is a very noticeable effect.  Another effect is in the deflection of the direction of the interstellar H-atom flow inside the heliosphere from the direction of the interstellar gas flow. The deflection can be measured by the deflection angle $\alpha_{defl}= arccos(V_{z,H}/V_H)$, where $V_{z,H}$ is the z component of the H atom velocity and  $V_H$ is the velocity. The deflection is associated with the change of the interstellar plasma streamlines around the heliopause due to the IsMF (Izmodenov et al. 2005b). Figures 7(C) and 7(I) show that the deflection angle between the models could be different by of the order of 1$^{\circ}$, which is small but might be important for the analyses of backscattered Lyman-alpha (see, e.g., Katushkina et al. 2015).

\section{Relevance of the model to observational data}

\begin{table}
	\tablename{ 1. Distances (in AU) to the TS and HP in the Upwind, {\it Voyager 1} (V1) and {\it Voyager 2} (V2) directions} \\
	\begin{tabular}{|c|c|c|c|c|c|c|c|c|c|}
		\hline Model &  TS, Upwind & TS, V1 & TS, V2  & HP, Upwind  &  HP,V1  & HP, V2   \\
		\hline 1    & 94.9 & 100.4& 89.7 & 153.8& 180.6 & 130.3 \\ 		
		\hline 2    & 83.7 & 88.5 & 81.0 & 153.0& 174.8 & 136.3 \\
		\hline 3    & 84.8 & 97.5 & 87.5 & 141.0& 163.5 & 138.5 \\
		\hline
	\end{tabular}
\end{table}

Before summarizing the results described in this paper, we have to give the reasons for our choice of the  proton and hydrogen number densities in the LISM because this was not done in subsection 2.1.
Although we did not aim in this paper to discuss the relevance of the model to all available observational diagnostics of the SW/LISM interaction region, we would like to note that the presented model has only four free parameters - interstellar proton and hydrogen number densities, magnitude of the IsMF, and the direction of the magnetic field in the ({\it BV}) plane. Other model parameters are not free since they are based on available observational information  discussed in detail in subsections 2.1 and 2.2.

In order to constrain the remaining free parameters, the following observational diagnostics can be used.

 \begin{enumerate}
   \item The distances of the heliospheric TS in the directions of the {\it Voyager 1} and {\it Voyager 2} spacecraft. They were determined by the spacecraft crossings as 83.7 and 94.1 AU  in 2004 August and in 2007 December, respectively.
   \item The number density of interstellar H atoms in the distant heliosphere (at 50-90 AU) of the order of 0.10$\pm$0.01 cm$^{-3}$. This value has been determined in different analyses of pick-up proton data (see, e.g., Geiss et al. 2006; Bzowski et al. 2008) and the SW deceleration at large heliospheric distances due to charge exchange with protons (e.g., Richardson et al. 2008).
   \item The difference of the interstellar H atom flow direction inside the heliosphere  compared with the direction of the pristine circumsolar LISM (Lallement et al. 2005, 2010).
   \item The SW plasma density and velocity data in the heliosheath from the Plasma/{\it Voyager 2} instrument.
   \item The distance to the heliopause in the direction of {\it Voyager 1} of 122 AU in  2012 August-September.
   \item The magnetic field strength and direction observed by MAG/{\it Voyager 1} after  2012 September (e.g.,  Burlaga \& Ness 2012).
   \item The proton number density inferred from analyses of the plasma wave instrument by Gurnett et al. (2013).
   \item It is widely believed in the community (e.g., Funsten et al. 2013; Schwadron \& McComas 2013; Heerikhuisen et al. 2014) that the center of the observed ribbon seen by IBEX is closely related to the direction of the IsMF and can serve as an additional constraint on the IsMF.
 \end{enumerate}

While this list of observational diagnostics is not complete (for example, energetic particle data should be added as well), it is quite evident that in the context of the current model paradigm
it is impossible to find the model parameters to reconcile all of the available diagnostics.

For example,  it is impossible to get the TS at 94 AU and the HP at 122 AU in {\it Voyager 1} in the frame of the global model with the same realistic boundary conditions. Solar cycle variations of the SW parameters can not resolve this problems as  is clearly seen from time-dependent simulations (see, e.g., Izmodenov et al. 2005, 2008), which show that the TS fluctuates with the solar cycle within $\pm$10 AU, while fluctuations of the HP are  $\pm$3-4 AU around its mean value.
In order to obtain  agreement between the observed distances to the TS and HP, Izmodenov et al. (2014) suggested  adding a physical effect such as thermal conduction, which would decrease the thermal pressure in the heliosheath.
Another example of diagnostic inconsistency is the direction of the IsMF. The direction of the {\it IBEX} ribbon center does not belong to the ({\it BV}) plane derived from backscatterred Lyman-alpha analyses by Lallement et al. (2010).
The detailed analyses of the inconsistencies is outside the scope of this paper and should be a subject of further study.

Since the model described in this paper has been used to study interstellar H atoms (e.g., Katushkina et al. 2015, this issue), we choose to fulfill the first three of the diagnostics listed above. The first is the distances to the TS, which provide the correct size of the heliosphere. The second and third are directly related to the interstellar atoms, and it makes sense to use this parameters for analyses of the interstellar H atom fluxes, backscattered solar Lyman-alpha and heliospheric absorption in Lyman-alpha. Since complete parametric studies are very numerically expensive for the considered 3D kinetic-MHD model, below we describe  how we obtained the interstellar proton and H atom densities used in this paper.

The interstellar proton and H-atom number densities of $\sim$0.06  and  $\sim$0.18 cm$^{-3}$ were established in the parametric study by Izmodenov et al. (2003). The axis-symmetric Baranov-Malama model was used in that study (see, also, Figure 1 in Izmodenov 2009). Izmodenov (2009) performed a new parametric study in the context of a 3D kinetic-MHD model (Model 1) and concluded that the model with the following set of model parameters gives  results agreeable with the observational diagnostics listed above: n$_{p,LISM}$ = 0.05$\pm$0.015 cm$^{-3}$, n$_{H,LISM}$ =0.18$\pm$0.018 cm$^{-3}$, B$_{LISM}$=2.5-3.5 $\mu$G, $\alpha_{LISM}$ = 15$^{\circ}$-30$^{\circ}$. Shortly afterward, Izmodenov et al. (2009) reported another possible solution with stronger IsMF: B$_{LISM}$=4.4$\mu$G, $\alpha_{LISM}$ = 20$^{\circ}$.

In the present study, we started with interstellar parameters obtained by Izmodenov et al. (2009). However, due to the presence of the HMF in the model, the TS is~ $\sim$10 AU closer to the Sun. In order to obtain a TS distance in the model comparable with  {\it Voyager's}  we have to reduce the interstellar pressure.
The reduction of the IsMF magnitude is not possible because a  smaller IsMF would lead to (1) less distortion of the heliopause and, therefore, smaller difference in the distances of the TS in the direction of {\it Voyager 1} and {\it 2}; and (2) smaller deflection angle $\alpha_{defl}$.
Therefore,  only the interstellar plasma pressure could be reduced. LISM velocity is not a free parameter because it is known from the {\it Ulysses}/Gas and {\it IBEX} observations. Therefore, to decrease the interstellar pressure, we have to decrease the interstellar proton number density n$_{p,LISM}$.
The decrease of n$_{p,LISM}$ would result in smaller filtration of H atoms  in the SW/LISM interaction region and, therefore, in a larger amount of neutrals inside the heliosphere. The latter would contradict the second of the observational constraints listed above.  Therefore, n$_{H,LISM}$ should be reduced as well. In this manner, we ended up with the values of
n$_{p,LISM}$ = 0.04 cm$^{-3}$, n$_{H,LISM}$ =0.14 cm$^{-3}$, B$_{LISM}$=4.4 $\mu$G, $\alpha_{LISM}$ = 20$^{\circ}$. With these parameters, our Model 3 satisfies (with some uncertainties) all of the observational constraints listed above. For the first and second diagnostics, see Table 1 and Figure 7, respectively. The third diagnostic - the deflection angle - was intensively studied in Katushkina et al. (2015).
 It is important to note that we did not aim to have exact distances in the {\it Voyager 1} and {\it Voyager 2} directions, since they would change with the solar cycle variations of the SW parameters (see, Izmodenov et al. 2008).

\section{Conclusions}

In this paper, we presented the results of our latest kinetic-(ideal)MHD model. The novel parts of the model are the HMF and the heliolatitudinal variations of the SW density and velocity.
The main results reported in this paper are as follow.

1. The HMF  results in the following:
\begin{itemize}
	\item closer heliocentric distance to the TS by~ $\sim$10 AU  compared to  models without the field;
	\item change of the heliopause shape when the heliocentric distances to the heliopause remain the same in the upwind part of the heliosphere and become smaller in the crosswind and tail parts;
	\item the increase of the HMF in the heliosheath toward the heliopause leads to the formation of the plasma depletion layer around the heliopause, similar to those observed in the magnetospheres of planets;
	\item the SW plasma flow in the inner heliosheath (and, in particular, at the heliopause) has general symmetry with respect to the solar equatorial plane, while some small asymmetries  connected with the distortion of the heliopause shape by the IsMF are observed;
	\item conversely, the interstellar plasma flow around the heliopause is oriented by the interstellar velocity and magnetic field directions, although asymmetries due to  the three dimensions of the heliopause exist as well.
\end{itemize}

2. The latitudinal variations of the SW dynamic pressure adopted in the model have maxima at middle heliolatitudes. This results in a more blunt heliopause when the heliopause becomes closer to the Sun in its upwind part and further away from the Sun on the flanks. The general orientations of the plasma flows inside and outside the heliopause remain the same - the SW flow is oriented around the solar equatorial plane, and the interstellar plasma flow is oriented around the plane determined by the interstellar velocity and magnetic field vectors.
We note here that in Model 3 there are no strong discontinuities in the magnetic field components through the heliopause in the {\it Voyager 1} direction. However, the detailed comparison of the model results with {\it Voyager} magnetic field data is not the subject of this paper and will be explored separately.

3. We have found that the model with the following parameters n$_{p,LISM}$ = 0.04 cm$^{-3}$, n$_{H,LISM}$ =0.14 cm$^{-3}$, B$_{LISM}$=4.4 $\mu$G, $\alpha_{LISM}$ = 20$^{\circ}$ provides the reasonable agreement with the distances of the heliospheric TS in the directions of the {\it Voyager 1} and {\it Voyager 2} spacecraft, the number density of interstellar H atoms in the distant heliosphere (at 50-90 AU) of the order of 0.10$\pm$0.01 cm$^{-3}$, and the difference of the interstellar H atom flow direction inside the heliosphere as compared with the direction of the pristine circumsolar LISM.

The model presented in this paper is not directly relevant to analyses of {\it IBEX}-Hi data because it does not consider the pick-up protons as a separate component, as  was done, for example, by Malama et al. (2006) for the axis-symmetric case or by Chalov et al. (2009) for the 3D case. Such a model is in progress and will be reported in the near future. At the same time, the model presented here is quite appropriate for analyzing of {\it IBEX}-Lo hydrogen data and has been used in the study presented in a companion paper by Katushkina et al. (2005, this issue).

{\bf Acknowledgements.} This paper is devoted to the memory of our senior colleague and Teacher - Y.G. Malama. We thank Olga Katushkina for her help with the preparation of this paper. Numerical calculations were performed using the supercomputers ``Lomonosov" and ``Chebyshev" of the Supercomputing Center of Lomonosov Moscow State University. The heliospheric part of the paper has been supported by RFBR grant 14-02-00746. The numerical modeling of the global heliosphere/astrosphere was performed under Russian Science Foundation grant 14-12-01096.

\end{document}